\documentclass{aastex63}
\usepackage{subfigure}
\usepackage{amsmath}
\usepackage{comment}
\usepackage{xcolor}
\usepackage{colortbl}
\usepackage{multirow}
\usepackage{placeins} %katia
\usepackage{subfiles} % Best loaded last in the preamble

\usepackage{xcolor}
\usepackage{soul}

\chardef\us=`\_

\received{\today}
\revised{\today}
\accepted{\today}
\submitjournal{ApJ}

\shorttitle{Geoffective events prediction and feature ranking through machine learning}
\shortauthors{Guastavino et al.}
\graphicspath{{./}{figures/}}
\begin{document}

\title{Forecasting Geoffective Events from Solar Wind Data and\\
Evaluating the Most Predictive Features through Machine Learning Approaches}

\correspondingauthor{Sabrina Guastavino}
\email{guastavino@dima.unige.it}

\author{Sabrina Guastavino}
\affiliation{MIDA, Dipartimento di Matematica, Università di Genova, via Dodecaneso 35 16146 Genova, Italy}
\affiliation{Istituto Nazionale di Astrofisica (INAF), Osservatorio Astrofisico di Torino}
\author{Katsiaryna Bahamazava}
\affiliation{Dipartimento di Scienze Matematiche Giuseppe Luigi Lagrange, Politecnico di Torino, Torino, Italy}
\author{Emma Perracchione}
\affiliation{Dipartimento di Scienze Matematiche Giuseppe Luigi Lagrange, Politecnico di Torino, Torino, Italy}
\author{Fabiana Camattari}
\affiliation{MIDA, Dipartimento di Matematica, Università di Genova, via Dodecaneso 35 16146 Genova, Italy}
\affiliation{Istituto Nazionale di Astrofisica (INAF), Osservatorio Astrofisico di Torino}
\author{Gianluca Audone}
\affiliation{Dipartimento di Scienze Matematiche Giuseppe Luigi Lagrange, Politecnico di Torino, Torino, Italy}
\author{Daniele Telloni}
\affiliation{Istituto Nazionale di Astrofisica (INAF), Osservatorio Astrofisico di Torino}
\author{Roberto Susino}
\affiliation{Istituto Nazionale di Astrofisica (INAF), Osservatorio Astrofisico di Torino}
\author{Gianalfredo Nicolini}
\affiliation{Istituto Nazionale di Astrofisica (INAF), Osservatorio Astrofisico di Torino}
\author{Silvano Fineschi}
\affiliation{Istituto Nazionale di Astrofisica (INAF), Osservatorio Astrofisico di Torino}
\author{Michele Piana}
\affiliation{MIDA, Dipartimento di Matematica, Università di Genova, via Dodecaneso 35 16146 Genova, Italy}
\affiliation{Istituto Nazionale di Astrofisica (INAF), Osservatorio Astrofisico di Torino}
\author{Anna Maria Massone}
\affiliation{MIDA, Dipartimento di Matematica, Università di Genova, via Dodecaneso 35 16146 Genova, Italy}



\begin{abstract}
This study addresses the prediction of geomagnetic disturbances by exploiting machine learning techniques. Specifically, the Long-Short Term Memory recurrent neural network, which is particularly suited for application over long time series, is employed in the analysis of in-situ measurements of solar wind plasma and magnetic field acquired over more than one solar cycle, from $2005$ to $2019$, at the Lagrangian point L$1$. The problem is approached as a binary classification aiming to predict one hour in advance a decrease in the SYM-H geomagnetic activity index below the threshold of $-50$ nT, which is generally regarded as indicative of magnetospheric perturbations. The strong class imbalance issue is tackled by using an appropriate loss function tailored to optimize appropriate skill scores in the training phase of the neural network. Beside classical skill scores, value-weighted skill scores are then employed to evaluate predictions,   suitable in the study of problems, such as the one faced here, characterized by strong temporal variability. For the first time, the content of magnetic helicity and energy carried by solar transients, associated with their detection and likelihood of geo-effectiveness, were considered as input features of the network architecture. Their predictive capabilities are demonstrated through a correlation-driven feature selection method to rank the most relevant characteristics involved in the neural network prediction model. The optimal performance of the adopted neural network in properly forecasting the onset of geomagnetic storms, which is a crucial point for giving real warnings in an operational setting, is finally showed.

%The results of our statistical analysis show that both parameters are good descriptors of the flaring proneness of an AR and possible tools for flare forecasting. 
\end{abstract}


\section{Introduction}
\label{sec:introduction}
Space Weather (SW) science studies how solar-terrestrial interactions affect the geospace environment \citep{2007LRSP....4....1P}; specifically, it involves predicting major solar disturbances that can pose significant risks to terrestrial facilities, with potential economic and security implications. The SW domain thus covers the various physical processes involved in the transfer of energy from solar wind and events to the Earth system, such as magnetic reconnection, the generation of ring currents in the terrestrial magnetosphere, and the interaction of solar charged particles with the Earth's atmosphere \citep{1931TeMAE..36...77C,1961PhRvL...6...47D,1966JGR....71..155F}. The most energetic events on the solar surface are flares \citep{piana2022hard}, intense electromagnetic emissions that can accelerate particles to relativistic velocities. They are often followed by massive eruptive events of chromospheric and coronal material, known as Coronal Mass Ejections \citep[CMEs;][]{2012LRSP....9....3W}, during which solar plasma and magnetic field are expelled into interplanetary space. Flares and CMEs, along with High-Speed Streams (HSS) of particles, and Corotating Interaction Regions (CIRs) and Heliospheric Current Sheet (HCS) crossings, can initiate geomagnetic storms that, when particularly intense, can have severe consequences for human activities and ground- and space-based infrastructures, such as telecommunications systems, satellite orbits, or power grids and pipelines.

Geomagnetic disturbances are usually detected by measuring perturbations in magnetospheric electric currents induced by solar storms. Specifically, the DST (Ap) geomagnetic index, derived from a network of ground-based magnetometer stations near the equator (at subauroral latitudes), is an estimate of the variations in the horizontal component of the ring currents circling the Earth in the equatorial plane (of the electric currents aligned with the Earth's magnetic field in the auroral ionosphere). DST and Ap thus provide an assessment of the severity of geomagnetic storms at low and high latitudes, respectively. The DST index (hourly acquired) or the equivalent SYM-H (acquired at one minute) are those most commonly deployed to reveal geomagnetic storms and measure their degree of geoeffectiveness. These are classified as moderate, intense or extreme depending on whether the DST index falls below thresholds of $-50$, $-100$ and $-250$ nT \citep{1998JGR...103..391C}.

Recently, machine learning and deep learning techniques are increasingly being employed in space weather applications \citep{2018SpWea..16....2C}, especially in the prediction of solar flares \citep{2015ApJ...798..135B,2016ApJ...829...89B,campi2019feature,Georgoulis_flarecast, guastavino2022implementation, guastavino2023operational,georgoulis2024prediction}, the onset of CMEs and their arrival time to Earth \citep{guastavino2023_physics,2023ApJ...948...78S}. Such studies are based on remote observations of the Sun and its atmosphere and, more specifically, in identifying CMEs and extracting their morphological/dynamic properties (such as angular width, velocity, and acceleration) from time sequences of coronagraphic white-light images, such as those from the Large Angle Spectroscopic Coronagraph (LASCO), as in \citet{pricopi2022predicting,Vourlidas2019}.

Less attention has been devoted, however, to predicting the degree of geoeffectiveness of solar events impinging on the Earth. In \citet{2023ApJ...952..111T} the problem was addressed using interplanetary plasma and magnetic field measurements (and, in particular, the intensity and the normal component to the ecliptic plane of the magnetic field vector, and the bulk speed, temperature, and density of the solar wind plasma) acquired in situ at the L$1$ Langragian point, $1.5$ million km from Earth in the Sunward direction, by resorting to the use of different types of neural networks. The recurrent architectures were found to be the best in predicting geomagnetic events associated with SYM-H values below $-50$ nT, achieving $94\%$ ($70\%$) accuracy when the prediction was made $1$ ($8$) hours in advance.

Similarly to \citet{2023ApJ...952..111T}, this paper aims to forecast the degree of severity of CME-induced geomagnetic storms, but explores and extends its predictive capabilities, using more refined and potentially better approaches. More specifically, as in the previous work, the problem was tackled as a binary classification with the goal of forecasting $1$-hour ahead a decrease in the SYM-H geomagnetic index below $-50$ nT, using the Long-Short Term Memory (LSTM) recurrent network \citep[which was shown to have the best performance in][]{2023ApJ...952..111T} applied to long time series of solar wind data. However, unlike \citet{2023ApJ...952..111T}, in which the strong data imbalance (SYM-H$<-50$ nT only $2\%$ of the time) potentially affecting the proper neural network training was addressed by exploiting data augmentation, in the present study this issue is faced by using an appropriate loss function designed to automatically optimize a suitable skill score for evaluating predictions in the case of highly imbalanced datasets \citep{marchetti2022score}. The advantage of this approach is to avoid data handling. The prediction performances are also evaluated against value-weighted skill scores \citep{guastavino2022bad} that are more suited for forecasting over time \citep{hu2022probabilistic,guastavino2022prediction} and will be shown to better evaluate performance in predicting the onset of geomagnetic storms. 
In addition, we have computed an estimate of the uncertainty of prediction by performing several runs of the neural network in the case of several generations of training, validation, and test sets. More importantly, in the present analysis, besides all interplanetary parameters (hereafter features) directly acquired by the instruments aboard the spacecraft, i.e., plasma and magnetic field measurements of the solar wind, some derived quantities, such as magnetic helicity \citep[which has been shown to be critical for properly detecting CMEs at L1, e.g.,][]{2019ApJ...885..120T} and solar wind transported energy \citep[which has been shown to play a crucial role in the magnetospheric response to solar drivers][]{2020ApJ...896..149T}, were also considered as inputs to the neural network. Finally, a correlation-driven feature selection method is here used to rank the most relevant, i.e., predictive, features involved in the neural network prediction model.

The layout of the paper is as follows: \S{} \ref{sec:dataset} presents the solar wind dataset used as input for the neural network model; \S{} \ref{sec:prediction_feature_ranking} describes the machine learning approach and correlation-driven feature ranking method; \S{} \ref{sec:results} shows the forecasting performances and the most predictive features. Our conclusions are offered in \S{} \ref{sec:conclusions}.

\section{Dataset}
\label{sec:dataset}
It is well known that a huge amount of data is required to train, validate and test neural networks. The dataset used in this paper consists of $7888319$ one-minute acquisitions related to multi-spacecraft interplanetary parameters acquired at L$1$ and geomagnetic activity indices measured on ground. The time period spans the years $2005$ to $2019$, thus encompassing the entire solar cycle $24$, and the descending (ascending) phase of the previous (subsequent) solar cycle $23$ ($25$). The Operating Missions as a Node on the Internet \citep[OMNI;][]{2005JGRA..110.2104K} database was used, which is intended specifically to study the effects of the solar wind variations on the Earth's magnetosphere. In fact, the interplanetary measurements are artificially time-shifted forward as if they had been acquired at the same location as the magnetospheric indices (in other words, the distance of $1.5$ million km separating L$1$ from Earth has been nullified by adding to the acquisition time of the heliospheric quantities the time it takes the solar wind to reach Earth). This allows the temporal correlations between solar and magnetospheric indices to be explored directly. Among all available interplanetary measurements, the following parameters were taken into account in the study: the magnitude and components of the magnetic field vector $\mathbf{B}$, the components of the flow velocity vector $\mathbf{V}$, along with the bulk speed $U$, density $\rho$, and temperature $T$ of the solar wind plasma. Some quantities derived from spacecraft measurements were also considered. These are the magnetic helicity $H_{m}$, the kinetic $E_{k}$ and magnetic energy $E_{m}$, and the total energy $E=E_{k}+E_{m}$. For a comprehensive discussion on how these quantities are estimated in the solar wind, the Reader is referred to \citet{2019ApJ...885..120T} and \citet{2020ApJ...896..149T}. Here it is only reminded that magnetic helicity is a measure of the degree of twisting of the magnetic field lines and is therefore a clear signature of the CME-embedded flux rope (which is a helical structure and therefore can be seen as a region of high magnetic helicity). It follows that $H_{m}$ is beneficial for a proper detection of CMEs. On the other hand, it appears evident that the more energetic the CME is (either due to its high velocity and/or the intensity of the associated magnetic field), the more severe the induced geomagnetic event is expected, since more energy will be transferred to the geospace. As a result, $E$ is a key parameter for correct prediction of geoeffectiveness of solar storms. As for the assessment of geomagnetic activity, the SYM-H index was instead employed.

As mentioned above, the goal of this study is to predict whether, based on a history of features over the past $24$ hours, in the next hour the SYM-H geomagnetic index will drop below the $-50$ nT threshold, customarily referred to for potential severe magnetospheric disturbances. From a purely computational perspective (i.e., to save machine resources), the time series are resampled at a one-hour resolution. This means that the total number of samples is thus reduced to just over $130000$. Since the study is approached as a binary prediction problem, samples corresponding to time periods when SYM-H $\lessgtr-50$ nT are labelled with $1$ or $0$, respectively. As already noted in \citet{2023ApJ...952..111T} and reported above, the dataset is highly imbalanced: only $2.53\%$ of total samples are labeled with $1$. The next section will present a suitable approach associated with the employment of an appropriate loss function to address this problem and enable proper training and validation of the neural network based on the use of suitable metrics for performance evaluation.

\section{Prediction and feature ranking}
\label{sec:prediction_feature_ranking}

\subsection{Assessment of results}
In order to compare the performances of machine learning methods for binary classification problems, as the prediction of geoffective events, the following points should be accounted for. First, the classification results should be evaluated by considering skill scores that are suitable for imbalanced data classification. Indeed, the percentage of geoffective events is really small, as already pointed out in Section \ref{sec:dataset}. Therefore, a chosen score needs to be capable of representing the performance of the classifier concerning the \textit{small} positive class. The classical skill scores are computed on the entries of the so-called confusione matrix, which is defined as
\begin{equation}\label{confusion-matrix}
\textrm{CM}= \left( \begin{array}{cc} 
\textrm{TN} & \textrm{FP} \\ 
\textrm{FN} & \textrm{TP}
\end{array}
\right),
\end{equation}
where the four entries are true positives (TPs), i.e. the number of samples labeled with $1$ and correctly predicted as positive; true negatives (TNs), i.e. the number of samples labeled with $0$ and correctly predicted as negative; false positives (FPs), i.e. the number of samples labeled with $0$ incorrectly predicted as positive; and false negatives (FNs), i.e. the number of samples labeled with $1$ and incorrectly predicted as negative. Among all possible skill scores, the True Skill Statistic (TSS) \citep{hanssen1965relationship}, is less sensible to the class-imbalance ratio \citep{bloomfield2012toward} than others, and therefore, it is particularly suitable for evaluation of imbalanced classification tasks. It is defined as the balance between the true positive and true negative rates (named also sensitivity and specificity, respectively), as follows \begin{equation}\label{eq:tss}
    \mathrm{TSS}(\mathrm{CM})=\frac{\mathrm{TP}}{\mathrm{TP}+\mathrm{FN}}+\frac{\mathrm{TN}}{\mathrm{FP}+\mathrm{TN}}-1=\mathrm{SENS}(\mathrm{CM})+\mathrm{SPEC}(\mathrm{CM})-1~,
\end{equation}
its values have range in the interval $[-1,1]$, and  the performance is optimal when $\mathrm{TSS}=1$, while $\mathrm{TSS}>0$ means that the rates of positive and negative events are mixed up. 

Second, the evaluation of binary predictions performed over time should take into account the forecast value, measured in terms of its usefulness to an operational setting to support the user while making a decision, as the importance in predicting the starting time of a geomagnetic storm. Value-weighted skill scores introduced by \cite{guastavino2022bad} that better account the intrinsic
dynamical nature of forecasting problems are defined on a value-weighted confusion matrix
that assigns different weights to FPs (denoted by wFPs) and FNs (denoted by wFNs) in such a way to account for the distribution of predictions along time with respect to the actual occurrences. By denoting the value-weighted confusion matrix as
\begin{equation}\label{confusion-matrix}
\textrm{wCM}= \left( \begin{array}{cc} 
\textrm{TN} & \textrm{wFP} \\ 
\textrm{wFN} & \textrm{TP}
\end{array}
\right),
\end{equation}
predictions are assessed by computing the value-weighted TSS (wTSS) defined as follows
\begin{equation}
    \textrm{wTSS} = \textrm{TSS}(\textrm{wCM}).
    \end{equation}

Finally, the splitting strategy between training, validation and test sets should take into account the rare-event nature of the problem, by maintaining uniformly the percentage of 1-labelled samples and the temporal distribution of events in order to not mix temporally samples in training, validation and test sets. Furthermore, the splitting strategy should be repeated several times in order to achieve some statistical significance. Therefore, many classification tests should be carried out by generating different triples of training, validation and test sets by maintaining the positive percentage of samples and  not drawing samples completely randomly between training and test with respect to time.

\subsection{Recurrent neural network}\label{sec:LSTM}
As neural network, we adopted the long short-term memory (LSTM) network \citep{Hochreiter1997LSTM}, which is the most widely adopted type of
recurrent neural network, able to process sequential data by solving the well-known short-term memory
problem of basic recurrent architectures. In the experiments we set the LSTM units equal to $72$ followed by a dense layer of $64$ neurons where the rectified linear unit (ReLU) activation funcion is adopted and 
a final dense sigmoid unit drives the output of the network to be in the interval $[0,1]$, in order to perform binary 
prediction.
The LSTM network is trained for $100$ epochs by taking batch size of $256$ samples, the \textit{Adam} optimizer \citep{Kingma14} is adopted for the optimization process with learning rate equal to $10^{-4}$.

In order to face the class-imbalance issue we adopted a suitable loss function which is designed to optimize an appropriate score. This strategy is introduced in \cite{marchetti2022score} and it is based on a probabilistic definition $\overline{\textrm{CM}}$ of the classical confusion matrix, depending on a chosen cumulative density function (cdf) on $[0,1]$ for the threshold parameter $\tau$ which separets $0$ and $1$.
%(in this application the cdf is chosen as the one related to the uniform distribution). The classical confusion matrix depends on a fixed threshold parameter $\tau \in (0,1)$, i.e., 
%\begin{equation}\label{confusion-matrix}
%\mathrm{CM}(\tau)= 
%\begin{pmatrix}
%\mathrm{TN}(\tau) & \mathrm{FP}(\tau) \\
%\mathrm{FN}(\tau) & \mathrm{TP}(\tau)
%\end{pmatrix}.
%\end{equation}
Let $(x_i,y_i)_{i=1}^n$ be input-label samples where $y_i\in\{0,1\}$ is the actual label associated to the sample $x_i$ and let $f(x_i)\in (0,1)$ is the probability outcome of the neural network $f$ on the sample $x_i$. We defined the probabilistic confusion matrix $\overline{\textrm{CM}}$ as
\begin{equation}
    \overline{\textrm{CM}} = \left( \begin{array}{cc} 
\overline{\textrm{TN}} & \overline{\textrm{FP}} \\ 
\overline{\textrm{FN}} & \overline{\textrm{TP}}
\end{array}
\right),
\end{equation}
characterized by the following entries 
\begin{equation}\label{eq:cm_elements}
\begin{array}{c}
    \overline{\textrm{TP}}=\sum_{i=1}^n y_i f(x_i),\quad
    \overline{\textrm{TN}}=\sum_{i=1}^n (1-y_i) (1-f(x_i))\\
    \overline{\textrm{FP}}=\sum_{i=1}^n (1-y_i) f(x_i),\quad 
    \overline{\textrm{FN}}=\sum_{i=1}^n y_i(1- f(x_i)).
\end{array}
\end{equation}
We chose the Score-Oriented Loss function (SOL) based on the TSS, which is defined as follows
\begin{equation}\label{TTS-loss}
\ell_{\mathrm{TSS}}:= - \textrm{TSS}(\overline{\textrm{CM}}).
\end{equation}
The main advantage consists in an automatic optimization of the desired skill score during the training phase without the need of a posteriori optimization of the threshold.

\subsection{Correlation-driven feature ranking}\label{sec:corr_feature_imp}

Feature ranking methods are commonly employed to identify a reduced subset of highly predictive features and to assess the relevance of physical attributes. One common technique consists in evaluating how each feature impacts the predictions according to some permutation importance score. Once a model is fitted via a training set, the permutation importance algorithm is implemented to evaluating how the accuracy of the prediction changes when a single feature is randomly shuffled in the validation dataset \citep{breiman2001random}. Indeed, when a feature is shuffled, its relevance increases as the accuracy score on the prediction decreases. Nevertheless, the permutation feature importance algorithm does not take into account the correlation between features. Therefore, we consider a correlation-driven feature importance method inspired by \cite{kaneko2022cross}, which includes the absolute correlation coefficients between features in the permutation process. Hence, in presence of strongly correlated features, such a method leads to more stable and reliable feature ranking schemes. In the following we summarize the steps of the correlation-driven feature importance algorithm:

\begin{itemize}
    \item Train a neural network $\hat{f}$.% on training data.
    \item Compute the prediction of the trained neural network $\hat{f}$ on the validation data, denoted as $\mathbf{Z}_{val}=\{(\mathbf{X}_{val},\mathbf{y}_{val})\}$ and compute the reference score $s_{val}$, i.e. the score computed between the prediction $\hat{f}(\mathbf{X}_{val})$ and $\mathbf{y}_{val}$.
    \item Calculate the correlation coefficients between all the features by following the procedure in \cite{kaneko2022cross}; in particular the correlation coefficient between two features is set to 0 if there is the possibility of chance correlation.
    \item For each feature $j$, i.e. the $j$-th column of validation data $\mathbf{X}_{val}$, and for each
repetition $k\in \{1,\dots, K\}$, randomly shuffle the $j$-th feature of the validation dataset $\mathbf{X}_{val}$, and for each feature $l\ne j$ for which the correlation value $c_{j,l}$ is higher than 0  randomly sample the column $l$ 
of $\mathbf{X}_{val}$ (without duplication with a probability of $c_{j,l}$) to
generate a corrupted version of the validation dataset, denoted as $\mathbf{X}^{j,k}_{val}$.
\item Compute the score $s_{j,k}$ between the prediction on the corrupted data $\mathbf{X}^{j,k}_{val}$, i.e. $\hat{f}(\mathbf{X}^{j,k}_{val})$, 
and $\mathbf{y}_{val}$.
\item Evaluate the importance $\mathcal{I}_j$ of the $j$-th feature by computing the difference between the reference score and the mean score obtained on the corrupted validation datasets, i.e.:
\begin{equation}
    \mathcal{I}_j = s_{val} - \frac{1}{K}\sum_{k=1}^K s_{j,k}.
\end{equation}
\end{itemize}
Such a procedure allows obtaining a feature ranking: the higher is $\mathcal{I}_j$, the higher is the contribution of the feature in the prediction. 

\section{Results}
\label{sec:results}
\begin{figure}
    \centering 
    \subfigure[{SYM-H not included in  features.}]{\includegraphics[width=0.7\textwidth]{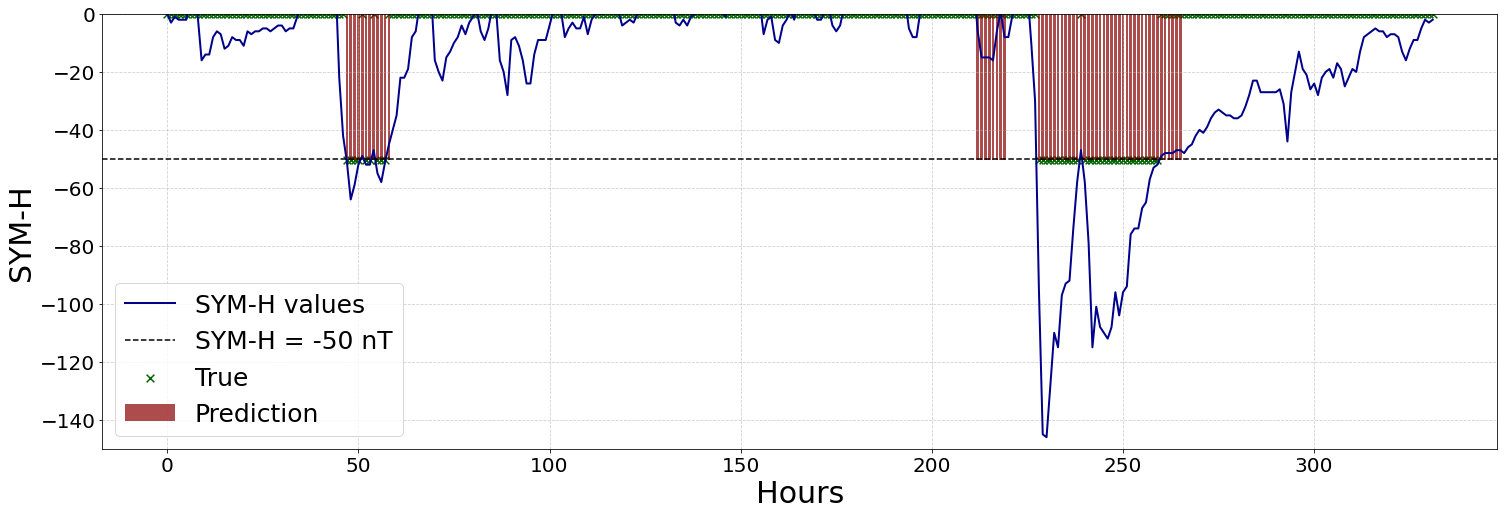}}\\ \subfigure[{SYM-H included in  features.}]{\includegraphics[width=0.7\textwidth]{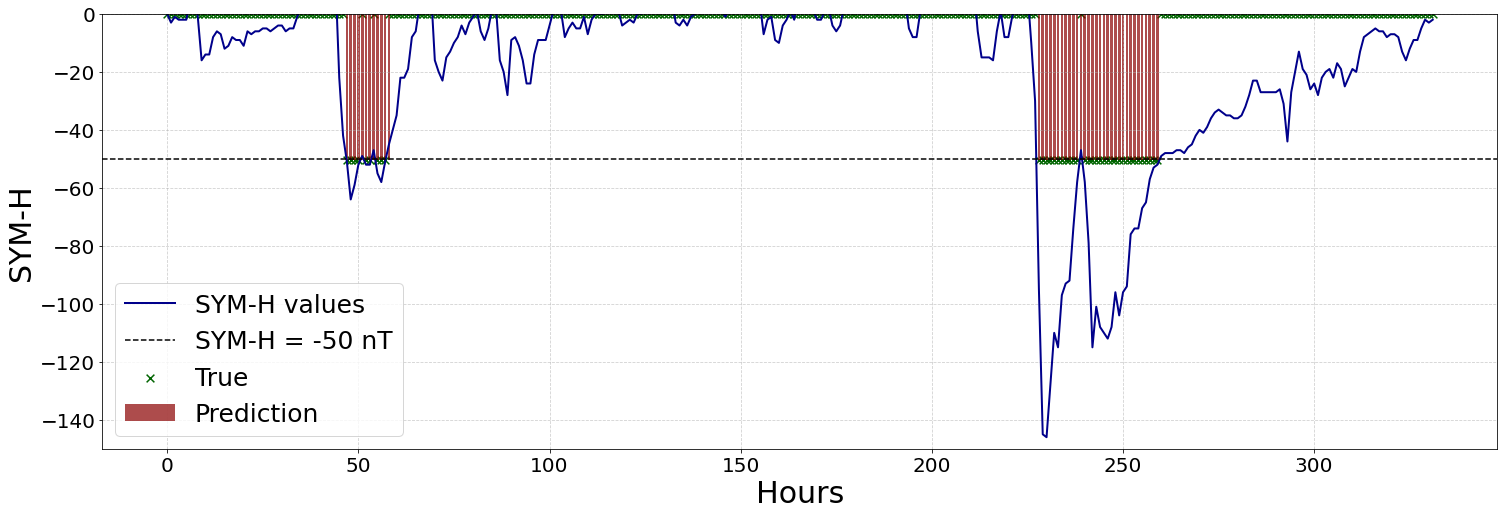}}
    \caption{Predictions over time on a temporal window of the test set of splitting 1: the top panel represents  the prediction when SYM-H is not included in the subset of features, whereas bottom panel represents the one when SYM-H is included between features.}\label{fig:pred_over_time_window1}
\end{figure}

\begin{figure}
    \centering
\subfigure[{SYM-H not included in  features}]{\includegraphics[width=0.7\textwidth]{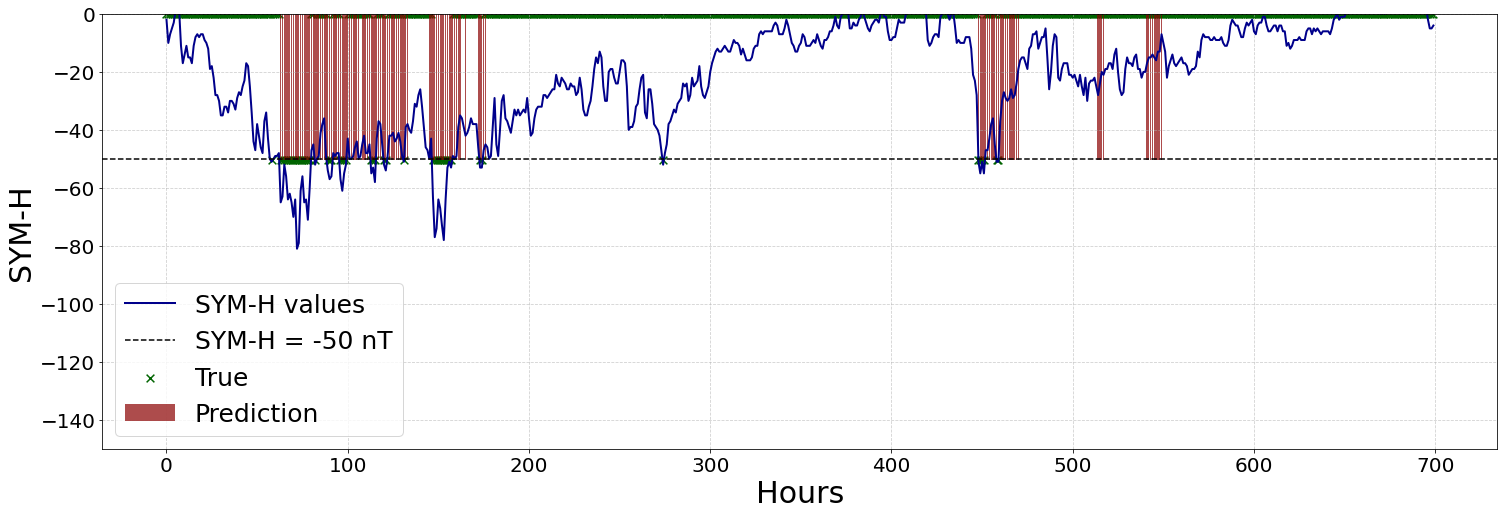}}\\
\subfigure[{SYM-H included in features}]{\includegraphics[width=0.7\textwidth]{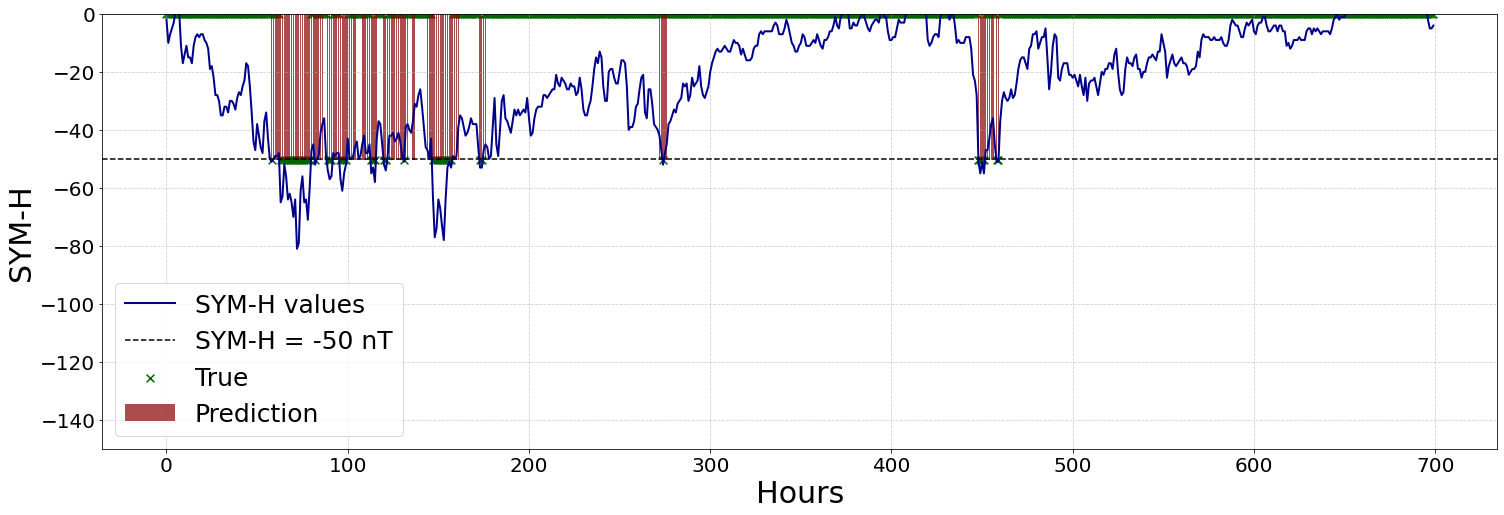}}
    \caption{Predictions over time on a temporal window of the test set of splitting 1: the top panel represents  the prediction when SYM-H is not included in the subset of features, whereas bottom panel represents the one when SYM-H is included between features.}\label{fig:pred_over_time_window2}
\end{figure}

The LSTM network described in  section \ref{sec:LSTM} has been applied to time series of features described in Section \ref{sec:dataset}. First a splitting strategy based on stratified K-fold splitting allows the generation of training, validation and test set that share the same rate of about $2.5\%$ of 1-labelled samples, leading to a training and validation sets which the sum represents about the $75\%$ of the total number of sample and a test set representing the remaining $25\%$ of samples. 
%#Train = 73 940
%#Val = 24 646
%#Test = 32 862, 
We investigated how the  prediction performances change if the SYM-H of the past 24-hour with cadence 1 hour is added as additional feature to the considered solar wind features, i.e. B, Bx, By, Bz (which represent the absolute value of the magnetic field vector $\mathbf{B}$ and the three components of $\mathbf{B}$), V, Vx, Vy, Vz (which represent the absolute value of the velocity vector $\mathbf{V}$ and the three components of $\mathbf{V}$), $\rho$ (density), T (temperature), $E_k$ (kinetic energy), $E_m$ (magnetic energy), $E$ (total energy) and $H_m$ (magnetic helicity). When the network is trained we applied the correlation-driven feature importance method described in section \ref{sec:corr_feature_imp}, in order to rank features accordingly to their predictive capabilities and we selected the first $10$ features which are associated to a meaningful importance value at least higher then $0.1$. In table \ref{tab:feature_ranking} we report the ranking of features with the associated importance value both in the case the SYM-H is included or not between features. In detail, we compute the confusion matrix CM and some common skill scores as the Heidke Skill Score (HSS) \citep{Heidke1926}, the sensitivity (SENS), the specificity (SPEC) defined in equation \eqref{eq:tss}, the F1 score (F1), which is the  harmonic mean of precision and sensitivity, and balanced accuracy (BA), which is the
arithmetic mean between sensitivity and specificity.
We noticed that the most predictive solar wind features are almost the same, both when SYM-H is included or not between features at least of permutation. In particular the energies, as the Total Energy is the third one in both rankings: this is coherent with the analysis conducted between the DST and energy in \cite{telloni2020magnetohydrodynamic}. Further, as we expected, features that are not correlated with the occurrence of geomagnetic storms have a negligible importance value less then 0.1.
In table \ref{tab:all_and_subset_features} we report the prediction performances on the test set when all the features were used and if the SYM-H is included in the list of features compared with the performance when only the subset of predictive features is considered. We noticed that the performances are higher when the subset of the most predictive features is adopted: this confirms the importance of a feature selection procedure also in the prediction phase. Finally in order to asses statistical robustness we reported results when 5 different splitting of training, validation and test sets are considered. In Table \ref{tab:results}, we reported the TSS and value-weighted TSS (wTSS) among the 5 generated test sets. We pointed out that when SYM-H is not used as feature the performances are low: this is due to the fact that when a geomagnetic storm is in place then, the SYM-H is below -50nT and such information helps the network to learn that if the previous hour the SYM-H is below -50nT the probability that in the next hour is below this threshold is high. Therefore, in figures \ref{fig:pred_over_time_window1} and \ref{fig:pred_over_time_window2} we report the distribution of predictions along time in correspondence of two different temporal windows which include some geomagnetic storms. When SYM-H is not used as additional feature we noticed that more false positives are present. However, from all panels it is clear that the starting time of the geomagnetic storm is well predicted even when SYM-H is not used as a feature: this confirms the importance of solar wind features that give information about the possible occurrence of a geomagnetic storm, whether caused by a coronal mass ejection or other events such as high speed streams.

\begin{table*}[ht]
\centering
\caption{Feature rankings with respect to the correlation-driven feature importance method shown in section \ref{sec:corr_feature_imp} in the case SYM-H is included between features or not.}\label{tab:feature_ranking}
% \resizebox{0.99\textwidth}{!}
{
\begin{tabular}{l  c | l   c }
\multicolumn{2}{c|}{Included SYM-H} &  \multicolumn{2}{c}{Not included SYM-H} \\
\hline
Feature & Importance value & Feature & Importance value \\
\hline
% \cline{1-5} \cline{6-9}\\[-8pt]
1) \textbf{SYM-H} & $0.5577$ & 1) \textbf{B} & $0.2936$  \\ 
2) \textbf{B} & $0.3392$ & 2) \textbf{Bz} & $0.2605$  \\ 
3) \textbf{Total Energy} & $0.3334$ & 3) \textbf{Total Energy} & $0.2175$  \\ 
4) \textbf{Vx} & $0.3256$ & 4) \textbf{Magnetic Helicity} & $0.2042$  \\ 
5) \textbf{V} & $0.3162$ & 5) \textbf{Vx} & $0.1902$  \\ 
6) \textbf{Temperature} & $0.287$ & 6) \textbf{Magnetic Energy} & $0.1901$  \\ 
7) \textbf{Magnetic Energy} & $0.2793$ & 7) \textbf{Temperature} & $0.18$  \\ 
8) \textbf{Magnetic Helicity} & $0.2628$ & 8) \textbf{V} & $0.1783$  \\ 
9) \textbf{Kinetic Energy} & $0.2427$ & 9) \textbf{Kinetic Energy} & $0.1604$  \\ 
10) \textbf{Bz} & $0.2303$ & 10) \textbf{Density} & $0.1451$  \\
11) Density & $0.2044$ & 11) Vy &  $< 0.1$  \\
12) Vy & $< 0.1$ & 12) By & $< 0.1$  \\
13) By & $< 0.1$ & 13) Bx & $< 0.1$  \\
14) Vz & $< 0.1$ & 14) Vz & $< 0.1$  \\
15) Bx & $< 0.1$& - & -  \\
\hline
\end{tabular}
}
\end{table*}

\begin{table*}[ht]
		\centering
		\caption{Comparison of the performances on the test set between using all features or the selected features with the correlation-driven feature importance method.
}\label{tab:all_and_subset_features}
\vspace{0.5cm}
% \resizebox{0.99\textwidth}{!}
{
\begin{tabular}{l | l l l l l l l l}
& \multicolumn{4}{c|}{Included SYM-H} & \multicolumn{4}{c|}{Not included SYM-H} \\
\cline{2-9}  \\[-12pt]
 & \multicolumn{2}{c|}{All features} &   \multicolumn{2}{c|}{Selected features} &   \multicolumn{2}{c|}{All features}  &   \multicolumn{2}{c|}{Selected features}\\
\multirow{2}{*}{
CM} &  TP=$629$  %\multicolumn{1}{c}{41} 
& \multicolumn{1}{c|}{FN=$204$} & TP=$707$ & \multicolumn{1}{c|}{FN=$124$} & TP=$495$ & \multicolumn{1}{c|}{FN=$336$} & TP=$550$ & \multicolumn{1}{c|}{FN=$281$}\\
& FP=$74$  %\multicolumn{1}{c}{41} 
& \multicolumn{1}{c|}{TN=$31599$} & FP=$430$ & \multicolumn{1}{c|}{TN=$615$} & FP=$36$ & \multicolumn{1}{c|}{TN=$668$} & FP=$430$ &  \multicolumn{1}{c|}{TN=$31599$} \\
\hline\\[-12pt]
TSS &  \multicolumn{2}{c|}{0.742
}  
 & \multicolumn{2}{c|}{0.8412}
 & \multicolumn{2}{c|}{0.5766
} & \multicolumn{2}{c|}{0.633
}\\ 
HSS &   \multicolumn{2}{c|}{0.655} &     \multicolumn{2}{c|}{0.7593} &     \multicolumn{2}{c|}{0.4963} 
 & \multicolumn{2}{c|}{0.4622} \\
%wFN  & \multicolumn{2}{c|}{14.25}  &     
%\multicolumn{2}{c|}{17.17}  &     
%\multicolumn{2}{c|}{25.92} 
 % \\
%wFP &    \multicolumn{2}{c|}{105.17} &     \multicolumn{2}{c|}{131.17} &     \multicolumn{2}{c|}{49.42}  \\
SENS 
  & \multicolumn{2}{c|}{0.755
}  
  & \multicolumn{2}{c|}{0.8508
} & \multicolumn{2}{c|}{0.5957}
  & \multicolumn{2}{c|}{0.6619
} 
 \\
SPEC &     \multicolumn{2}{c|}{0.987}  & \multicolumn{2}{c|}{0.9904} & \multicolumn{2}{c|}{0.9809} & \multicolumn{2}{c|}{0.9715}   \\
F1 &     \multicolumn{2}{c|}{0.665}  & \multicolumn{2}{c|}{0.766} & \multicolumn{2}{c|}{0.511} & \multicolumn{2}{c|}{0.48} \\
BA &     \multicolumn{2}{c|}{0.871}  & \multicolumn{2}{c|}{0.9206} & \multicolumn{2}{c|}{0.7883} & \multicolumn{2}{c|}{0.8267} \\
\hline
\end{tabular}
}
\end{table*}

%\begin{table*}[ht]
%\centering
%\caption{Mean and standard deviation values of the scores on the 5 realizations of the test sets}\label{tab:results}
% %\resizebox{0.99\textwidth}{!}
%{
%\begin{tabular}{ c| c  c c c  | c c c c  }
%\multirow{1}{*}{Score} & \multicolumn{4}{c|}{Included SYM-H} &  \multicolumn{4}{c}{Not included SYM-H} \\
% \cline{2-9}
%& mean & std & min & max  & mean & std & min & max \\
%\hline
%% \cline{1-5} \cline{6-9}\\[-8pt]
%TSS & $0.8742$ & $0.028$ &$0.9082$ & $0.8442$  & $0.7154$ & $0.076$ & $0.8203$ & $0.6296$ \\ 
%wTSS & $0.8164$ & $0.044$ & $0.8973$ $0.7838$ & $0.6166$  & $0.6022$ & $0.109$ & $0.7597$ & $0.4757$  \\  
%\hline
%\end{tabular}
%}
%\end{table*}

\begin{table*}[ht]
\centering
\caption{Performances on 5 test sets in case the SYM-H is included or not in the subset of features.}\label{tab:results}
% \resizebox{0.99\textwidth}{!}
{
\begin{tabular}{c | c  c  | c c }
\multirow{2}{*}{Splitting} & \multicolumn{2}{c|}{TSS} &  \multicolumn{2}{c}{wTSS} \\
 \cline{2-5}
& Included SYM-H & Not included SYM-H & Included SYM-H & Not included SYM-H  \\
\hline
% \cline{1-5} \cline{6-9}\\[-8pt]
1 & $0.9073$ & $0.82029$ &$0.8964$ & $0.7597$   \\ 
2 & $0.8654$ & $0.7146$ & $0.8503$ & $0.6166$   \\ 
3 & $0.9082$ & $0.7791$ & $0.8973$ & $0.6743$  \\ 
4 & $0.8442$ & $0.6296$ & $0.7838$ & $0.4844$ \\ 
5 & $0.846$ & $0.6334$ & $0.8164$ & $0.4757$  
 \\ 
\hline
Mean $\pm$ std & $0.8742\pm 0.028$ & $0.7154\pm 0.076$ & $0.8164\pm 0.044$ & $0.6022\pm 0.109$  
 \\ 
\hline
\end{tabular}
}
\end{table*}

\section{Conclusions}
\label{sec:conclusions}
In this work, the performance of the LSTM recurrent neural architecture in providing one-hour advance alerts of critical SYM-H values less than $-50$ nT and thus indicative of geomagnetic disturbances was estimated. It turns out that its predictive capability reaches up to $91\%$ and $82\%$ in TSS in the case of considering or not SYM-H itself as an input feature of the neural network. It appears evident that in the former case the neural network is able to predict the activity status of the Earth's magnetosphere even after the onset of the geomagnetic storm, that is, to estimate its recovery phase toward a quiet condition. This results in higher performance, as it is easier to also forecast the duration of the recovery phase of a solar storm rather than just its onset, which is precisely what LSTM does when not having the value of SYM-H as input. From an operational point of view, it is also worth mentioning that the solar wind data used in the analysis were artificially time-shifted as if they had been acquired at the same location as the geomagnetic indices. However, if the tool worked in real time in forecasting the value of SYM-H in the hour after the acquisition of solar measurements at L$1$, the time required for the solar plasma to reach Earth, i.e., $30$ minutes ($60$ minutes) for a bulk speed of $800$ ($400$) km s$^{-1}$, would also have to be accounted for. This means that the prediction of geomagnetic disturbances could be provided with $91\%$ efficiency up to $2$ hours before the onset of the geomagnetic storm, ranking the present approach among those currently available that provide at the same time earlier and more accurate warnings.

\FloatBarrier  % This will prevent figures from floating past this point

\section*{Acknowledgement}
\label{sec:acknowledge}
SG was supported by the Programma Operativo Nazionale (PON) “Ricerca e Innovazione” 2014–2020; AMM, EP, RS, DT, GN acknowledge the support of the Fondazione Compagnia di San Paolo within the framework of the Artificial Intelligence Call for Proposals, AIxtreme project (ID Rol: 71708). AMM is also grateful to the HORIZON Europe ARCAFF Project, Grant No. 101082164.

%SG and FM acknowledge the financial support of the Programma Operativo Nazionale (PON) “Ricerca e Innovazione” 2014–2020. VC was supported by the Universit\`a di Genova within the Bando per l’incentivazione alla progettazione europea (BIPE) - Mission 1 “Promoting Competitiveness” 2020.

%{\bf{(AMM SICURAMENTE RINGRAZIA ARCAFF. CHI ALTRI LO RINGRAZIA? DIREI SOLO AMM. FORSE AMM RS E DT POSSONO RINGRAZIARE ANCHE AIXTREME?)}}

\bibliography{refs}{}
\bibliographystyle{aasjournal}

\end{document}